\documentclass[11pt]{article}
\PassOptionsToPackage{hyperfootnotes=false}{hyperref}
\usepackage[preprint]{acl}
\usepackage{times}
\usepackage{inconsolata}
\usepackage{latexsym}
\usepackage[T1]{fontenc}
\usepackage[utf8]{inputenc}
\usepackage{amsmath}
\usepackage{amssymb}
\usepackage{array}
\usepackage{multirow}
\usepackage{makecell}
\usepackage{bbding}

\usepackage{diagbox}
\usepackage{mathrsfs}
\usepackage{color}
\usepackage{colortbl}
\usepackage{enumitem}
\usepackage{algorithm}
\usepackage{algpseudocode}
\usepackage{tcolorbox}
\usepackage{needspace}
\tcbuselibrary{breakable,listings}
\newtcblisting{promptbox}[1]{
  listing only, breakable,
  colback=black!3, colframe=black!40, boxrule=0.5pt, arc=2.5pt,
  left=4pt, right=4pt, top=3pt, bottom=3pt,
  title={#1}, fonttitle=\bfseries\footnotesize, coltitle=black,
  colbacktitle=black!8,
  listing options={basicstyle=\ttfamily\footnotesize,
    breaklines=true, columns=fullflexible, keepspaces=true,
    breakindent=0pt, postbreak={}}}
\usepackage{graphicx}
\usepackage{wrapfig}
\usepackage{caption}
\definecolor{skyblue}{RGB}{135,206,235}
\definecolor{lightblue}{RGB}{235,245,255}
\usepackage{booktabs}
\usepackage{microtype}
\newcommand{\pace}{PACE-Bench}
\title{When Coding Agents Outpace Review: Evaluating PRD-Driven Development in Mature Enterprise Repositories}
\author{Tianqi Liu$^{1,*}$\thanks{Work done during an internship at Alibaba Taobao \& Tmall Group.} \quad Yandong Guan$^{2,*}$ \quad
  Songze Li$^{1}$ \quad Wenkai Wang$^{1}$ \\
  Gang Yu$^{2,\dagger}$ \quad Chengfei Lv$^{2,\dagger}$ \\
  $^1$Zhejiang University \qquad $^2$Alibaba Taobao \& Tmall Group \\
  \texttt{tqliu@zju.edu.cn} \\
  $^*$Equal contribution \qquad $^\dagger$Corresponding authors }
\hypersetup{pdfauthor={Tianqi Liu, Yandong Guan, Songze Li, Wenkai Wang, Gang Yu, Chengfei Lv},pdftitle={When Coding Agents Outpace Review: Evaluating PRD-Driven Development in Mature Enterprise Repositories}}
\begin{document}
\setcounter{footnote}{2}

\maketitle
\begin{abstract}
Coding agents increasingly undertake development tasks in mature
enterprise repositories, yet existing benchmarks offer limited insight
into whether they fulfill complete Product Requirement Documents (PRDs)
and where delivery breaks down. Issue resolution and development from
scratch leave incremental PRD fulfillment underexplored, while isolated
tests provide incomplete evidence of business and integration requirements.
We introduce \pace{}, a benchmark of 43 enterprise tasks with 750
expert-audited Code Points. It combines requirement reconstruction,
criterion construction, and evidence-based evaluation to measure PRD
fulfillment and connect delivery gaps to agent behavior.
Across 11 model--scaffold configurations, most scores fall between 52.1
and 78.1; GPT-5.6-Sol achieves the highest complete-delivery rate of 14.0\%.
Incomplete implementation and missing integration are the principal
bottlenecks: producing individual components does not ensure a complete
delivery. Evaluation achieves 95.2\% agreement with expert consensus,
with lower case-level variation across judging records than repeated
executions. Beyond model comparison, the benchmark provides actionable
feedback for harness refinement: diagnosis guides targeted improvements,
and successive harness releases raise mean performance from 59.4 to 63.0
using the same model and task suite. Code is available at
\url{https://github.com/TianqiLiu1/PACE-Bench}.
\end{abstract}

\section{Introduction}
\label{sec:intro}
LLM-based coding agents increasingly support enterprise software
development \citep{murali2024codecompose,paradis2025how,xu2025theagentcompany}.
In practice, work is organized around \textbf{Product Requirement
Documents (PRDs)} that combine business rules, boundary conditions,
technical constraints, and acceptance criteria \citep{wiegers2013software}.
Fulfilling a PRD in a mature repository requires more than producing a
new component: agents must locate relevant code and coordinate changes
across modules, services, and configuration while preserving established
behavior. They must fit existing interfaces and architectural constraints \citep{hopkins2008brownfield}.

\begin{table*}[t]
\centering\small
\newcommand{\tabyes}{\textcolor[rgb]{0.20,0.46,0.40}{\Checkmark}}
\newcommand{\tabno}{\textcolor[rgb]{0.66,0.37,0.34}{\XSolid}}
\setlength{\tabcolsep}{3pt}
\renewcommand{\arraystretch}{1.18}
\begin{tabular*}{\textwidth}{@{\extracolsep{\fill}}llrccl@{}}
\toprule
Benchmark & Task input & \makecell{New / edited code\\(LOC / files)} & \makecell{Existing\\codebase} & \makecell{Generated\\acceptance criteria} & Evaluation method \\
\midrule
SWE-bench~\citeyearpar{jimenez2024swebench} & Issue description & $\sim$15 / 1 & \tabyes & \tabno & Unit + regression tests \\
DevEval~\citeyearpar{li2025devevalfull} & PRD & $\sim$413 / $\sim$4.4$^{\dagger}$ & \tabno & \tabno & Unit + acceptance tests \\
FEA-Bench~\citeyearpar{li2025feabench} & Feature request & 128.5 / 2.6$^{\dagger}$ & \tabyes & \tabno & Unit tests \\
SWE-bench Pro~\citeyearpar{deng2025swebenchpro} & Issue description & 107.4 / 4.1$^{\dagger}$ & \tabyes & \tabno & Unit + regression tests \\
PRDBench~\citeyearpar{fu2026prdbench} & PRD & $1{,}279^{\ddagger}$ / n/r & \tabno & \tabyes & \makecell[l]{Unit / CLI / file tests\\with an agent judge} \\
FeatureBench~\citeyearpar{zhou2026featurebench} & Feature request & 790.2 / 15.7$^{\dagger}$ & \tabyes & \tabno & Unit + regression tests \\
Harvest~\citeyearpar{jha2026reap} & Production prompt & 39 / 2 & \tabyes & \tabno & Fail to pass tests \\
\midrule
\textbf{\pace{}} & \textbf{PRD} & \textbf{1,465 / 15} & \tabyes & \tabyes & \makecell[l]{\textbf{Layered semantic}\\\textbf{Code Point scoring}} \\
\bottomrule
\end{tabular*}
\caption{Comparison of task inputs, code scale, and evaluation design. Generated acceptance criteria are synthesized requirements for judging candidate implementations. Code size uses medians; $\dagger$ marks means, and $\ddagger$ denotes an annotation scaffold rather than patch edits. Measurement scopes appear in Appendix~\ref{app:benchmark-comparison}.}
\label{tab:benchmark-position}
\end{table*}

Table~\ref{tab:benchmark-position} shows that generated acceptance
criteria are available for PRD-driven development from scratch, while
the listed suites for existing repositories mainly rely on executable
test oracles. \pace{} brings generated, auditable acceptance contracts
to PRD-driven brownfield tasks. This enables criterion-level assessment
of substantial incremental changes---a median of 1,465 changed lines
across 15 files---and traces incomplete delivery to specific business
and integration requirements.

The first challenge is \textbf{reliable acceptance judgment}.
Passing unit tests does not establish fulfillment of all PRD conditions,
business conventions, and integration requirements. End-to-end tests
provide broader evidence but require costly reconstruction of services
and data. An LLM judge can assess semantic fulfillment, yet its decisions
may depend on how criteria and evidence are presented
\citep{zheng2023judging,wang2024fairevaluators}. A useful evaluator must
therefore cover the intended requirements and produce stable judgments
when the requirements and candidate implementation remain unchanged.

The second challenge is \textbf{understanding delivery beyond a score}.
The same score can reflect different unmet requirements and different
execution failures. An absent implementation, a disconnected component,
and an unverified change call for different interventions. Outcome
attribution identifies what was not delivered; execution trajectories
show how the agent retrieved, edited, and verified code
\citep{deshpande2025trail,zhang2025which}; process observations and quality
indicators reveal aspects of the run and artifact that functional scores
do not capture. These complementary views are needed to form and check
explanations before choosing a model or changing its harness.

\Needspace{3\baselineskip}
We introduce \textbf{\pace{}} (\textbf{P}RD-\textbf{A}nchored
\textbf{C}oding-agent \textbf{E}valuation),
which turns enterprise development records into audited acceptance
criteria, scores candidate patches against frozen contracts, and links
delivery gaps to execution evidence and quality evidence.

Construction consolidates task documents and later user clarifications into a standalone PRD that preserves the final scope without making names introduced only by generated code mandatory. Independent PRD-only and reference-diff-only passes then extract required behavior and observed changes. Reconciliation filters incidental edits and unsupported inferences; automated refinement splits the survivors into atomic Code Points, and experts review their grounding and coverage. Each point's contract, allowed alternatives, and weight are frozen before agent evaluation.\begin{figure*}[t]
\centering
\includegraphics[width=\textwidth]{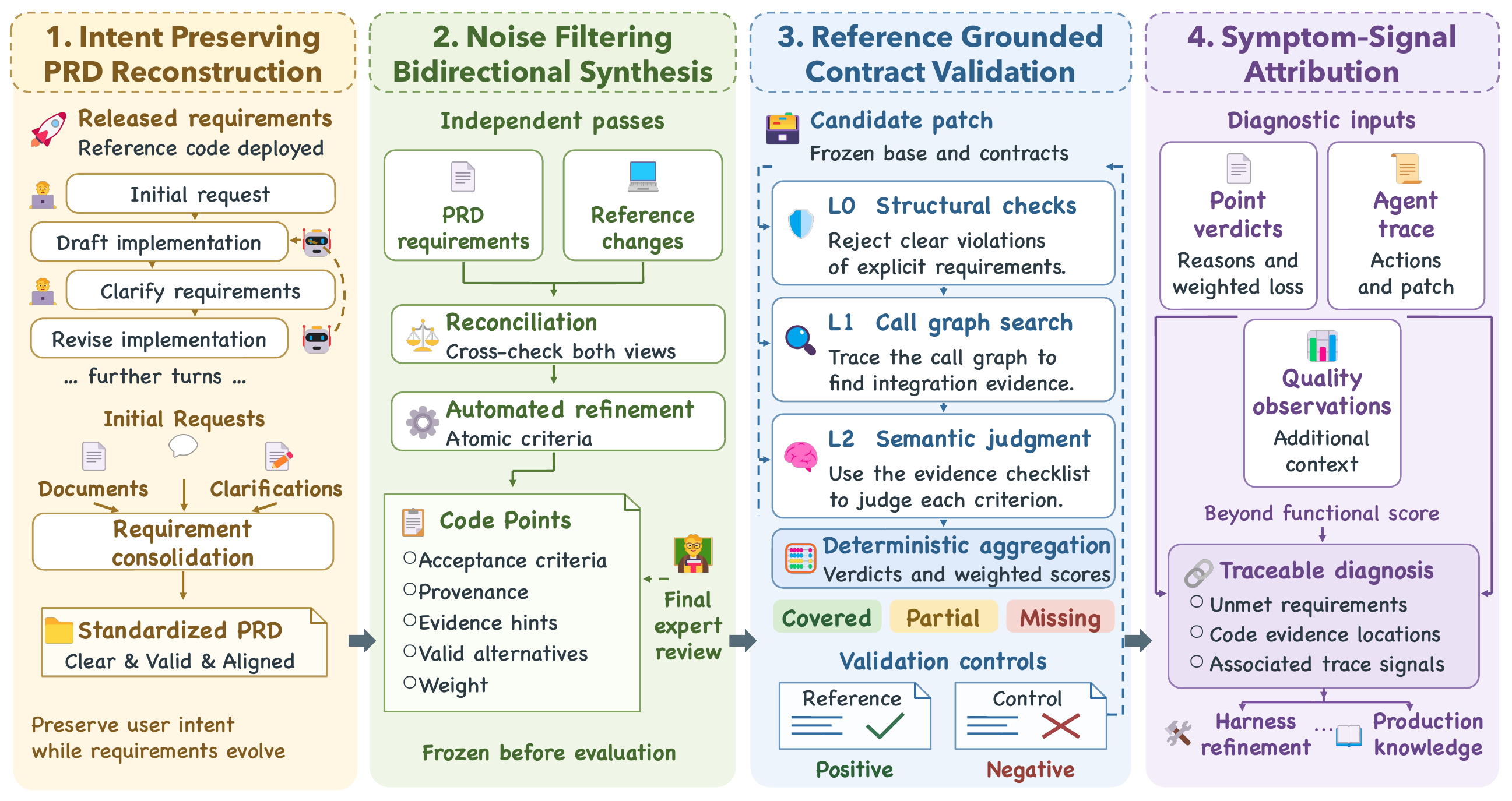}
\caption{\pace{} turns production requirements into frozen acceptance criteria, evidence-grounded judgments, and traceable diagnoses for agent improvement.}
\label{fig:overview}
\end{figure*}

Scoring applies the frozen contract in three stages. L0 structural checks reject conclusive violations of explicit constraints; L1 uses bounded call-graph search to gather candidate code and integration context; L2 judges each criterion with cited evidence while admitting equivalent implementations. Deterministic rules fold these judgments into Covered, Partial, or Missing and weighted scores. Accepted reference patches and four negative controls probe the scoring boundary.

Finally, weighted acceptance losses identify unmet PRD conditions and pair them with retrieval, editing, and verification signals from agent trajectories. Quality observations add context without altering functional credit. This linked evidence supports traceable diagnosis and informs harness revisions checked on the same frozen suite.

\Needspace{5\baselineskip}
Our contributions are threefold:
\begin{itemize}[leftmargin=*,itemsep=1pt,topsep=2pt]
\item We introduce \pace{} to evaluate coding agents on
\textbf{PRD-driven incremental development in enterprise brownfield
repositories}, pairing 43 production tasks with 750 expert-audited
Code Points that measure full-request fulfillment.
\item We develop noise-filtering bidirectional criterion synthesis and
a verifiability-stratified evaluator. Frozen contracts, localized evidence,
and deterministic aggregation yield auditable judgments and trace-linked
diagnoses.
\item We benchmark 11 model--scaffold configurations, finding only
\textbf{14.0\% complete PRD delivery} at best. Our evaluator achieves
\textbf{95.2\% agreement} with independent expert consensus; further
analysis identifies delivery bottlenecks and informs harness refinement.
\end{itemize}

\section{Benchmark Construction}
\label{sec:suite}
Figure~\ref{fig:overview} summarizes the pipeline.
\paragraph{Task definition.}
Given a standardized PRD $R^{\star}$ and base repository $C_0$, an agent
produces a patch $\Delta_a$. Frozen Code Points $G$ specify acceptance
conditions and weights. Agents receive the same PRD and base, without reference patches, source conversations, or Code Points.

\subsection{Intent Preserving PRD Reconstruction}
\label{sec:reconstruction}
A production record combines a multi-turn user--agent conversation,
base repository, accepted diff, and task metadata. Documents and user
clarifications define the final scope. Reconstruction consolidates them
into a standalone PRD that preserves evolving intent and requests work
from the frozen base. Names introduced solely by generated code cannot
become mandatory constraints; metadata checks alignment without creating
requirements. All agents therefore receive one consolidated final request (Appendix~
\ref{app:prd-protocol}).

\subsection{Noise Filtering Bidirectional Synthesis}
\label{sec:synthesis}
The PRD pass extracts behavior without seeing implementation; the diff
pass records changes without inferring business purpose. Reconciliation
cross-checks these views. Refinement makes criteria atomic, removes noise, and records evidence hints and allowed alternatives:
\begin{equation}
\begin{aligned}
\widetilde G &= \mathrm{Reconcile}\big(E_R(R^{\star}),E_{\Delta}(\Delta_r)\big),\\
G &= \mathrm{Review}\big(\mathrm{Refine}(\widetilde G)\big).
\end{aligned}
\end{equation}
Here $E_R$ and $E_{\Delta}$ are independent extraction passes.
Incidental reference changes do not become obligations. PRD-only findings
are flagged for expert review; reference-linked candidates are not
assumed to be fully satisfied.

\paragraph{Acceptance contracts and expert review.}
\label{sec:governance}
Each Code Point records behavior, provenance, retrieval hints,
permitted alternatives, and weight. Symbolic contracts additionally
specify required structures or interface values. A retrieval hint alone does not mandate an implementation.
\textbf{Weights reflect verifiability and business importance, not code volume.}
An initial weight distinguishes explicit, directly checkable contracts
from criteria requiring broader semantic judgment. Business priority
then adjusts the weight to reflect the consequence of an unmet requirement.

The platform team automates screening and validation. Business-domain
experts join only for final selection and audit, checking requirement
grounding, ambiguity, duplicates, and missing key acceptance items for
all 750 retained criteria. A separate, non-overlapping group of technical
annotators manually re-evaluates all 750 patch verdicts (Section~\ref{sec:human}).
The two reviews address criterion validity and scoring validity,
respectively; Appendix~\ref{app:curation} details the construction stages.

\begin{figure}[t]
\centering
\includegraphics[width=\columnwidth]{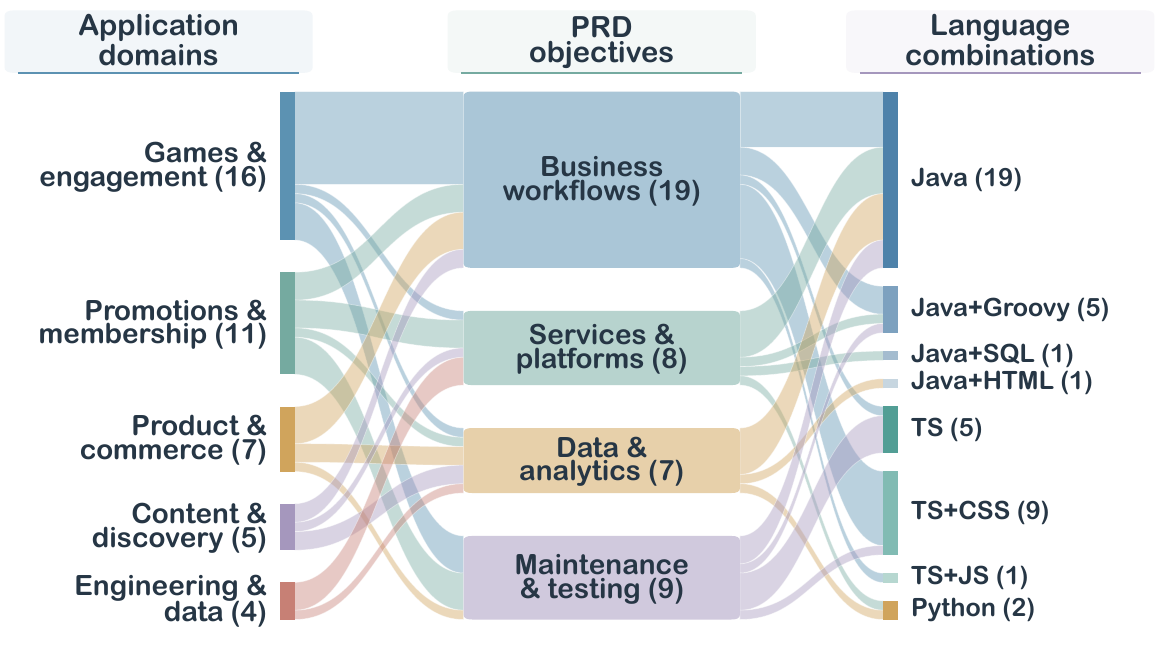}
\caption{Application domains, PRD objectives, and language combinations across the 43 production cases.}
\label{fig:dataset-profile}
\end{figure}

\paragraph{The resulting suite.}
\label{sec:stats}
The 43 production tasks contain 750 Code Points, with median reference
changes of 1,465 lines across 15 files. Figure~\ref{fig:dataset-profile}
shows the domains, objectives, and languages; Figure~\ref{fig:complexity}
characterizes repository and task scale. Context and change scope are distinct:
a compact patch can depend on a large codebase. Appendix~\ref{app:benchmark-comparison}
details these measures.

\section{Benchmark Evaluation Protocol}
\label{sec:eval}
\subsection{Reference Grounded Contract Validation}
\label{sec:judgment}
Each Code Point defines required behavior and permitted implementations.
The protocol separates three questions: whether a violation is conclusive,
where the candidate provides relevant evidence, and whether that evidence
establishes fulfillment. A shared contract aligns these stages without treating retrieval as proof of fulfillment.

\paragraph{L0: Deterministic Contract Validation.}
L0 checks explicit acceptance constraints that can be resolved
mechanically. It closes a point as Missing only when the inspected scope
proves a violation; otherwise it defers to L1. It does not award Covered.
This conservative boundary prevents missing names or unfamiliar
implementations from becoming unsupported rejection decisions.

\paragraph{L1: Call Graph Evidence Localization.}
L1 connects each requirement to the candidate's changed code and its
integration context. It retrieves both the implementation and the
connections needed to make the behavior reachable. Unchanged base code
explains those connections but earns no independent acceptance credit.
A bounded expansion can resolve a specific evidence gap. The goal is an
inspectable evidence package, not a functional verdict inferred from
identifier presence. Appendix~
\ref{app:cascade} specifies the stopping rules.

\paragraph{L2: Constrained Semantic Judgment.}
L2 assesses the fixed acceptance conditions against the localized
evidence, admitting \textbf{equivalent implementations}. Each condition
receives a cited judgment; the judge cannot introduce new requirements.
Partial means real but incomplete fulfillment, rather than uncertainty
about unseen code. Criterion-level judgments make the fulfilled behavior
and remaining gap explicit, while fixed aggregation determines the
point verdict and score. Prompt and response details appear in
Appendix~\ref{app:prompt-judge}.

\paragraph{Deterministic aggregation.}
A hard miss gives Missing; a partial entry or soft miss gives Partial;
all criteria satisfied gives Covered. For resolved verdicts,
\begin{equation}
S_c=100\frac{\sum_{i\in G_c}w_i\phi(y_i)}{\sum_{i\in G_c}w_i},
\label{eq:score}
\end{equation}
where $\phi$ maps Covered, Partial, and Missing to $1$, $0.5$, and $0$.
Unresolved evidence blocks a final score rather than removing weight.
Frozen weights encode verifiability and business importance; system scores average
cases equally. \textbf{Complete delivery} requires $S_c\geq95$ and zero Missing.
Success rate is the fraction of cases meeting this criterion.

\begin{figure}[t]
  \centering
  \includegraphics[width=\columnwidth]{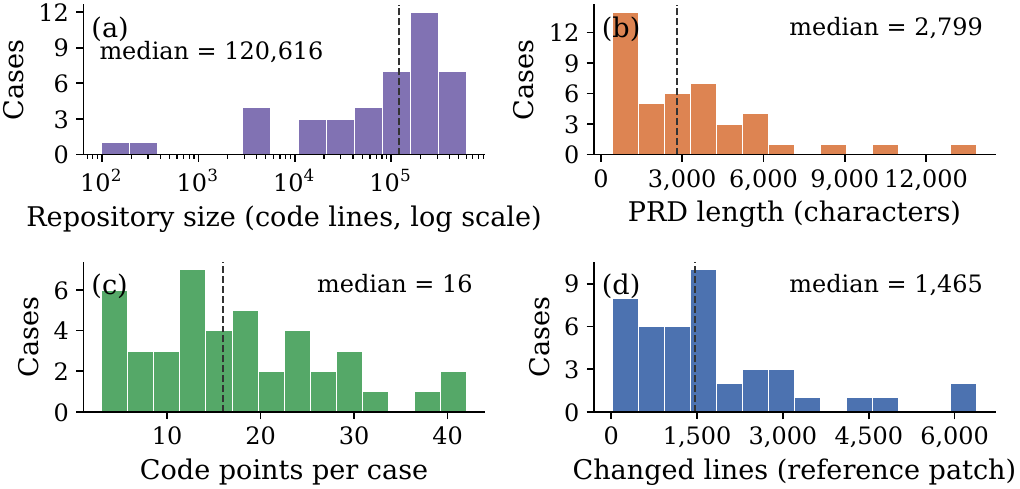}
  \caption{Task-scale distributions: repository size, PRD length, acceptance scope, and reference edits. Dashed lines mark medians.}
  \label{fig:complexity}
\end{figure}

\subsection{Symptom--Signal Attribution}
\label{sec:attribution}
Attribution keeps three views distinct. \textbf{Outcome symptoms} map
cited non-Covered reasoning to contract mismatch, integration gap,
test-delivery gap, or incomplete implementation; weighted loss
$w_i(1-\phi(y_i))$ links each symptom to the score. \textbf{Process
signals} record retrieval, editing, and verification in agent traces.
\textbf{Quality observations} describe the delivered artifact without
changing functional credit. Outcome gaps and process signals guide trace inspection. 

\section{Benchmark Experiments}
\label{sec:experiments}

\subsection{Experimental Setup}
\label{sec:setup}
We compare 11 model--scaffold configurations. Eight primary models are evaluated on the same 43 tasks using a common
OpenHands scaffold \citep{wang2025openhands}: GPT-5.6-Sol, GLM-5.2,
Qwen3.7-Max, Kimi-K2.6, Opus-4.8, Gemini-3.1-Pro-Preview,
DeepSeek-V4-Pro, and MiniMax-M2.7. Native-scaffold configurations pair
GPT-5.6-Sol with Codex and Qwen3.7-Max with QwenCode. The main table also includes Qwen3.8-27B. Agents receive the
standardized PRD and base repository, without the reference patch
or acceptance criteria.

All scores use fixed Code Points and weights. We report case means
with standard errors; task outcomes receive equal weight across the suite.
Each configuration in Table~\ref{tab:models} is evaluated over the 43-case suite,
with a verdict for every one of the 750 Code Points.
Appendix~\ref{app:config} gives the evaluation configuration.
Cross-model diagnostic analyses use the eight primary common-scaffold
models. Native configurations compare complete agent systems.

\subsection{RQ1: Model and Scaffold Performance}
\label{sec:models}
We first compare how much of each request agents fulfill and how often
they satisfy the complete acceptance contract.

\Needspace{4\baselineskip}
\noindent\textbf{Obs.1: Partial fulfillment and complete delivery
reveal complementary model and scaffold capabilities.}

Table~\ref{tab:models} reports mean functional score and
complete-delivery success.
Among the eight primary common-scaffold models, mean scores span 26.0
points, from 52.1 to 78.1. GPT-5.6-Sol leads both measures, completing \textbf{6 of 43 tasks}.
Partial fulfillment and complete delivery remain distinct: models with
similar mean scores can differ in the number of fully satisfied requests.

\begin{table}[htp]

\centering\small
\setlength{\tabcolsep}{2pt}
\begin{tabular*}{\columnwidth}{@{\extracolsep{\fill}}>{\raggedright\arraybackslash}p{.48\columnwidth}rr@{}}
\toprule
Model & \makecell{Score\\(mean $\pm$ SE)} & \makecell{Success\\rate} \\
\midrule
\multicolumn{3}{l}{\emph{Unified minimal scaffold}} \\
GPT-5.6-Sol               & 78.08 $\pm$ 2.71 & 14.0\% \\
GLM-5.2                   & 68.80 $\pm$ 3.72 & 9.3\% \\
Qwen3.7-Max               & 66.70 $\pm$ 3.45 & 7.0\% \\
Kimi-K2.6                 & 63.09 $\pm$ 3.79 & 4.7\% \\
Opus-4.8                  & 61.68 $\pm$ 4.39 & 7.0\% \\
Gemini-3.1-Pro-Preview    & 61.30 $\pm$ 3.84 & 2.3\% \\
DeepSeek-V4-Pro           & 61.00 $\pm$ 3.65 & 9.3\% \\
MiniMax-M2.7              & 52.13 $\pm$ 3.64 & 2.3\% \\
Qwen3.8-27B  & 12.25 $\pm$ 3.12 & 0.0\% \\
\midrule
\multicolumn{3}{l}{\emph{Native scaffolds}} \\
GPT-5.6-Sol (Codex)       & 76.78 $\pm$ 2.81 & 14.0\% \\
Qwen3.7-Max (QwenCode)    & 65.69 $\pm$ 3.83 & 11.6\% \\
\bottomrule
\end{tabular*}
\caption{Functional score (mean $\pm$ SE) and complete-delivery success over 43 cases.}
\label{tab:models}
\end{table}

\paragraph{Evaluating the complete agent system.}
Codex and the unified GPT-5.6-Sol setup
both complete 14.0\% of cases; their mean scores are 76.8 and 78.1,
respectively. QwenCode raises Qwen3.7-Max's complete-delivery rate from
7.0\% to 11.6\%, with a mean score of 65.7 versus 66.7 under the unified
scaffold. These paired
configurations illustrate that scaffold effects differ across systems.
For Qwen3.7-Max, the native scaffold completes five cases rather than
three, despite a 1.01-point lower mean. A scaffold can change which requirements are completed together
without increasing average fulfillment.

\begin{figure}[t]
  \centering
  \includegraphics[width=\columnwidth]{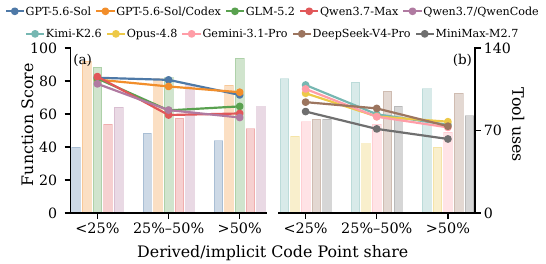}
  
  \caption{Performance by derived/implicit Code Point share. Lines show scores; pale bars show mean tool use.}
  \label{fig:difficulty}
\end{figure}

\paragraph{Task difficulty.}
Observed difficulty in this suite is more closely associated with
\textbf{inferring required behavior} than with simple task size.
Figure~\ref{fig:difficulty} groups cases by the share of derived or
implicit Code Points: $<25\%$, $25\%$--$50\%$, and $>50\%$.
A larger share means more acceptance conditions must be inferred from
PRD intent rather than read directly from its wording.

The groups score 75.0, 61.6, and 56.8 across the eight models, an
18.2-point difference between the outer groups. All eight models have
negative correlations with this share. By comparison, Code Point count,
changed lines, changed files, and PRD length have weak correlations with
mean score ($|\rho|\leq.083$). This pattern highlights interpreting and
connecting requirements as a more informative difficulty factor than
volume alone. Appendix~\ref{app:difficulty-details} reports correlations,
uncertainty intervals, and the exploratory comparison across descriptors.

\subsection{RQ2: Diagnosing Delivery Gaps}
\label{sec:findings}
We read each delivery through three linked views: weighted acceptance
loss identifies the unmet requirement, trajectory signals locate relevant
agent behavior, and auxiliary quality observations describe the artifact.

\Needspace{4\baselineskip}
\noindent\textbf{Obs.2: Acceptance-level evidence identifies
incomplete implementation and integration as targets for agent improvement.}

\begin{figure}[htp]
  \centering
  \includegraphics[width=\columnwidth]{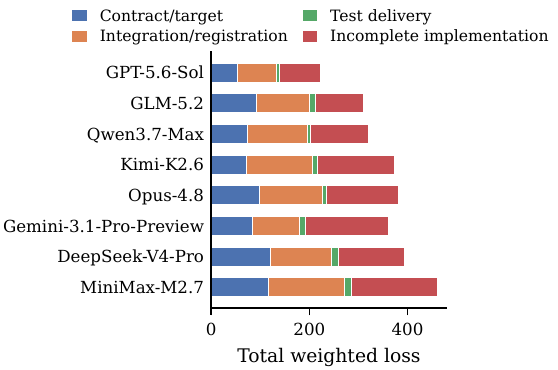}
  
  \caption{Weighted acceptance loss across 43 cases, decomposed into four outcome symptoms.}
  \label{fig:symptoms}
\end{figure}

Figure~\ref{fig:symptoms} decomposes weighted loss into four mutually
exclusive descriptive symptoms. Incomplete implementation is the
largest category for six of the eight models, while integration or
registration is largest for GLM-5.2 and Qwen3.7-Max. Integration remains
a substantial source of lost credit across all models. Together, these
two symptoms account for \textbf{72.0\% of pooled weighted loss}, ranging
from 65.8\% to 78.1\% within individual models. The categories summarize
cited acceptance evidence; trajectory inspection connects them to
execution decisions. An incomplete handler and an unregistered handler
need different interventions: implementing behavior versus updating a
connection.

\begin{figure*}[htp]
  \centering
  \includegraphics[width=\textwidth]{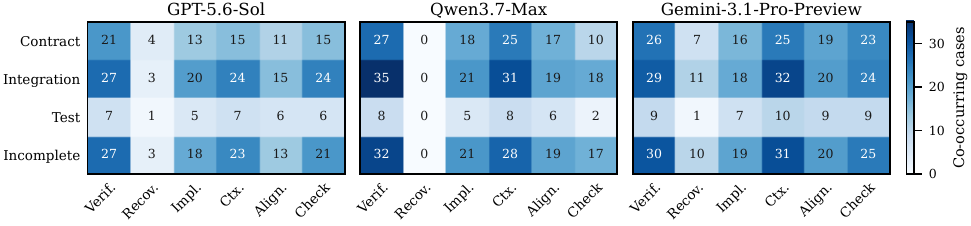}
  
  \caption{Case-level co-occurrence of outcome symptoms (rows) and process signals (columns) for three models, using a shared color scale.}
  \label{fig:cooccur}
\end{figure*}

Trajectory signals add the execution view (Figure~\ref{fig:cooccur}).
Verification gaps are the most frequent companions of acceptance
misses, followed by expected files that were never retrieved. Qwen
concentrates on verification gaps, while Gemini also shows an
error-recovery signal. These trace-linked co-occurrences direct
investigation toward context selection, entry-point wiring, and
post-edit checks; engineers inspect the corresponding events to decide
which intervention is warranted. Section~\ref{sec:deployment} shows a
retrieval-policy diagnosis under a fixed context budget.

A reviewer follows each loss to the cited declaration or edge, then
checks its retrieval, edits, and verification in the trace.
Table~\ref{tab:observations} shows quality and process observations
for each configuration. Naming-contract scores cluster at 89.9--95.5,
whereas test-delivery scores span 22.3--55.3 on applicable cases;
verification commands also vary, from 10.4 to 20.3 for the two
GPT-5.6-Sol scaffolds. These observations guide inspection without
changing functional scores.

The three views answer different questions. An outcome symptom names
the violated acceptance contract and accounts for its weighted loss.
A trajectory signal points to an event worth inspecting, such as a
missing retrieval or absent post-edit verification; its co-occurrence
with a symptom does not by itself prove the event caused the miss.
Quality observations describe the delivered artifact's conventions,
architecture, naming, and test coverage even when these properties are
not direct PRD requirements. Keeping them separate prevents a cleanly
named but disconnected component from gaining functional credit, while
still giving engineers concrete leads for review and harness changes.

\begin{table*}[t]
\centering\footnotesize
\setlength{\tabcolsep}{2.5pt}
\begin{tabular*}{\textwidth}{@{\extracolsep{\fill}}lrrrrrrrr@{}}
\toprule
& \multicolumn{4}{c}{Code quality (0--100)} & \multicolumn{4}{c}{Execution process} \\
\cmidrule(lr){2-5}\cmidrule(l){6-9}
Model & Conv. & Arch. & Name & Test & Verify & Self & Post & Tools \\
\midrule
\multicolumn{9}{l}{\emph{Unified minimal scaffold}} \\
GPT-5.6-Sol & 58.2 & 68.4 & 94.3 & 47.2 & 10.4 & 8.4 & 5 & 62.0 \\
GLM-5.2 & 57.7 & 66.9 & 93.2 & 51.1 & 5.9 & 4.0 & 11 & 122.8 \\
Qwen3.7-Max & 52.5 & 62.8 & 94.0 & 45.4 & 4.7 & 4.3 & 4 & 75.9 \\
Kimi-K2.6 & 57.4 & 65.1 & 95.5 & 43.2 & 8.0 & 6.9 & 10 & 110.1 \\
Opus-4.8 & 54.9 & 65.9 & 89.9 & 49.6 & 4.8 & 3.1 & 14 & 59.8 \\
Gemini-3.1-Pro-Preview & 48.9 & 63.2 & 93.0 & 41.5 & 4.6 & 3.4 & 3 & 73.9 \\
DeepSeek-V4-Pro & 53.1 & 64.0 & 91.7 & 22.3 & 4.4 & 4.3 & 0 & 95.8 \\
MiniMax-M2.7 & 51.1 & 63.6 & 92.4 & 36.2 & 5.8 & 5.3 & 10 & 84.8 \\
\midrule
\multicolumn{9}{l}{\emph{Native scaffolds}} \\
GPT-5.6-Sol (Codex) & 59.1 & 78.2 & 93.2 & 55.3 & 20.3 & 14.2 & 16 & 116.7 \\
Qwen3.7-Max (QwenCode) & 59.5 & 67.9 & 95.5 & 38.3 & 1.7 & 0.7 & 17 & 97.2 \\
\bottomrule
\end{tabular*}
\caption{Auxiliary code-quality and process observations. Verify, Self, and Tools are mean counts per case; Post counts cases with successful verification after the last edit (of 43). Test scores use the nine cases requiring tests. None enters the functional score.}
\label{tab:observations}
\end{table*}

\subsection{RQ3: Reliability of Acceptance Evaluation}
\label{sec:suite-validity}
We test whether evaluation distinguishes relevant implementations from
negative controls and agrees with the fixed expert-consensus reference.

\Needspace{4\baselineskip}
\noindent\textbf{Obs.3: Positive and negative controls distinguish
requirement fulfillment from superficial code matches.}

With the PRD, base, criteria, and weights fixed, we compare accepted
references with four negative controls: (A) no changes, (B) unrelated
same-repository edits, (C) another case's patch, and (D) expected names
or locations with empty class or method bodies. We verify exact-base applicability before scoring.
Table~\ref{tab:negative-controls} summarizes the results;
Appendix~\ref{app:controls} details construction.

\begin{table}[t]
\centering\small
\begin{tabular*}{\columnwidth}{@{\extracolsep{\fill}}lr@{}}
\toprule
Control & Score / 100 \\
\midrule
\multicolumn{2}{@{}l}{\emph{Positive control}} \\
Accepted reference patch & \textbf{98.35} \\
\midrule
\multicolumn{2}{@{}l}{\emph{Negative controls}} \\
A. Empty patch & \textbf{0.00} \\
B. Same-repository unrelated patch & \textbf{0.00} \\
C. Applicable cross-case patch & \textbf{0.00} \\
D. Identifier-only patch & \textbf{4.03} \\
\bottomrule
\end{tabular*}
\caption{Reference and negative-control scores.}
\label{tab:negative-controls}
\end{table}

\paragraph{Positive control.}
Reference patches score 98.35 on average, with a median of 100.00.
Business-domain experts checked every non-Covered point against the
PRD and reference implementation. They confirmed omitted implementation
details that the PRD explicitly required, and accepted the corresponding
Partial or Missing verdict. The shortfall therefore reflects genuine
acceptance gaps, rather than a requirement that references score perfectly.

\paragraph{Negative controls.}
A--C earn no credit on applicable patches. Cross-case patches that
cannot apply to the target base are excluded, rather than scored zero.
D receives no Covered verdicts. Its three Partials have expected
declarations or locations but empty bodies: the semantic judge grants
structural credit without recognizing complete behavioral fulfillment.

\paragraph{Expert agreement and component contributions.}
\label{sec:human}
The component study tests how contract grounding, evidence localization,
and aggregation contribute to agreement under identical patches.

\Needspace{4\baselineskip}
\noindent\textbf{Obs.4: Contract grounding, evidence localization,
and checklist aggregation jointly support agreement with expert consensus.}

We compare seven judging methods
under identical candidate patches and the frozen 750-point consensus
reference (Table~\ref{tab:agreement}). Three platform-side technical
annotators manually re-evaluated all 750 Qwen patch verdicts to form
the consensus reference; none participated in the business-domain
criterion audit. Appendix~\ref{app:annotation} describes the review protocol. We report exact agreement, linearly weighted Cohen's $\kappa$,
and severe Covered--Missing inversions.

The three ablations change one component at a time. Without contract
grounding, the judge receives the description without its detailed
acceptance contract. Without tiered judgment, the complete diff replaces
localized evidence. Without checklist aggregation, the judge emits a
verdict directly. Baselines use a vanilla judge, a vanilla judge supplied
with the contract, and an agent-as-a-judge that searches the repository
autonomously \citep{zhuge2024agentjudge}.

The full pipeline achieves \textbf{95.2\% agreement} and $\kappa=.945$, with
no severe inversions. Removing contract grounding reduces $\kappa$
to .595, removing checklist aggregation to .677, and removing tiered
judgment to .843. The autonomous judge reaches 78.0\% agreement and
$\kappa=.726$. Together, these comparisons support combining a precise
acceptance boundary with localized evidence and deterministic verdict
aggregation. The component comparison fixes candidate patches and consensus
labels across methods and is reported separately from the
model-delivery study in RQ1.

The full pipeline exceeds the strongest baseline by 17.2 agreement
points. The contract fixes what must be established, localization fixes
the inspected evidence, and aggregation fixes the verdict rule.
\begin{table}[t]
\caption{Expert-consensus agreement, grouped into the full pipeline, component ablations, and judge baselines.}
\label{tab:agreement}
\centering\small
\setlength{\tabcolsep}{2pt}
\renewcommand{\arraystretch}{1.08}
\begin{tabular*}{\columnwidth}{@{\extracolsep{\fill}}>{\raggedright\arraybackslash}p{.46\columnwidth}rrr@{}}
\toprule
Method & Agr. $\uparrow$ & \makecell{Weighted\\$\kappa$ $\uparrow$} & \makecell{Severe\\dis. $\downarrow$} \\
\midrule
\textbf{Full pipeline} & \textbf{95.2\%} & \textbf{0.945} & \textbf{0.0\%} \\
\midrule
w/o contract grounding & 70.3\% & 0.595 & 9.7\% \\
w/o tiered judgment & 87.7\% & 0.843 & 1.5\% \\
w/o checklist aggregation & 76.3\% & 0.677 & 5.9\% \\
\midrule
Vanilla LLM judge & 77.6\% & 0.686 & 4.3\% \\
Vanilla LLM with contract & 76.0\% & 0.652 & 8.3\% \\
Agent-as-a-judge with contract & 78.0\% & 0.726 & 1.5\% \\
\bottomrule
\end{tabular*}
\end{table}

\subsection{RQ4: Delivery Consistency}
\label{sec:stability}
We compare repeated deliveries with judging records for fixed patches
to distinguish execution variability from evaluation variability.

\Needspace{4\baselineskip}
\noindent\textbf{Obs.5: Judging is more stable than execution.}

For Qwen3.7-Max, the three executions score 66.70, 65.44, and 65.37.
Three judging records for the fixed Run~1 patches score 66.70, 66.47,
and 66.40, with complete delivery in 3, 2, and 3 of 43 cases,
respectively (Table~\ref{tab:stability}). We average each case's
three-record sample standard deviation and range over the 43 cases:
mean case SD is 8.77 for execution and 2.36 for judging; mean case
range is 16.34 and 4.46. Execution variation is therefore about
\textbf{3.7 times larger} by both measures. This per-case comparison
holds candidate patches fixed for judging, so the smaller variation
does not rely on similar aggregate means alone.

\begin{figure}[t]
\centering
\includegraphics[width=0.98\columnwidth]{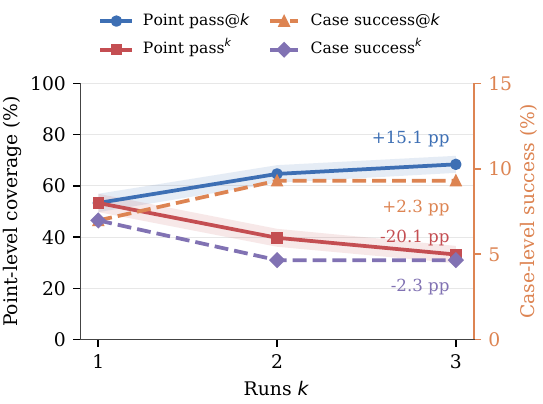}
\caption{Qwen3.7-Max repetition: any-run versus every-run success. Annotations show changes in percentage points.}
\label{fig:passk}
\end{figure}

Figure~\ref{fig:passk} distinguishes acceptance at least once from
acceptance in every run. Point-level pass@$3$ reaches 68.4\%, whereas
pass$^3$ is 33.2\%. Relative to one run, these represent changes of
$+15.1$ and $-20.1$ percentage points. Case-level any-run and every-run
success are 9.3\% and 4.7\%. This distinction shows why aggregate
means alone do not describe delivery reliability: similar run means
can conceal different subsets of fulfilled requirements.

The smaller fixed-patch variation supports \textbf{judge repeatability}.
Similar functional means still hide delivery differences, motivating
\textbf{beyond-functional analysis} of execution signals and quality
observations.

\subsection{RQ5: Does Static Knowledge Improve Implementation?}
\label{sec:knowledge}
We select 28 cases with available task-relevant repository knowledge,
covering 525 Code Points, to study static knowledge injection with
Qwen3.7-Max. For each selected case, we compare a no-knowledge input
with three fixed knowledge inputs: repository locators, task-matched
repository knowledge, and shared engineering documentation. All four
inputs are evaluated on this subset in a paired comparison.

\Needspace{4\baselineskip}
\noindent\textbf{Obs.6: Static knowledge injection yields no consistent gains.}

\begin{table}[t]
\centering\small
\setlength{\tabcolsep}{2.3pt}
\begin{tabular*}{\columnwidth}{@{\extracolsep{\fill}}>{\raggedright\arraybackslash}p{.44\columnwidth}rr@{}}
\toprule
Knowledge input & \makecell{Mean score\\$\pm$ SE} & $\Delta$ \\
\midrule
No knowledge & $58.54\pm5.11$ & --- \\
Locators only & $57.80\pm4.84$ & $-0.74$ \\
Repository-matched knowledge & $59.23\pm4.77$ & $+0.69$ \\
Shared long context & $57.03\pm5.22$ & $-1.51$ \\
\bottomrule
\end{tabular*}
\caption{Paired knowledge comparison on 28 cases selected for available repository knowledge.}
\label{tab:knowledge}
\end{table}

Table~\ref{tab:knowledge} shows changes from $-1.51$ to $+0.69$
points. Knowledge availability alone therefore
does not reliably translate into effective implementation in this setting.
Each reported delta compares a knowledge input with the no-knowledge
input on the same selected cases.
Moreover, a small change in mean score can mask different requirements
being completed across cases. The result supports a limited conclusion
about these fixed inputs and this suite, rather than a general claim
that repository knowledge cannot help coding agents.
The distinction motivates examining how context enters the execution
process, rather than treating additional documentation as sufficient.
Appendix~\ref{app:knowledge} describes construction, contents, and fallback.

\subsection{RQ6: Can Diagnosis Guide Harness Improvement?}
\label{sec:deployment}
We applied the evaluation--diagnosis--repair workflow to our coding-agent
platform's harness and compared successive releases on the fixed suite.

\Needspace{4\baselineskip}
\noindent\textbf{Obs.7: Trace-linked diagnosis identifies actionable harness changes.}

\begin{figure}[t]
  \centering
  \includegraphics[width=\columnwidth]{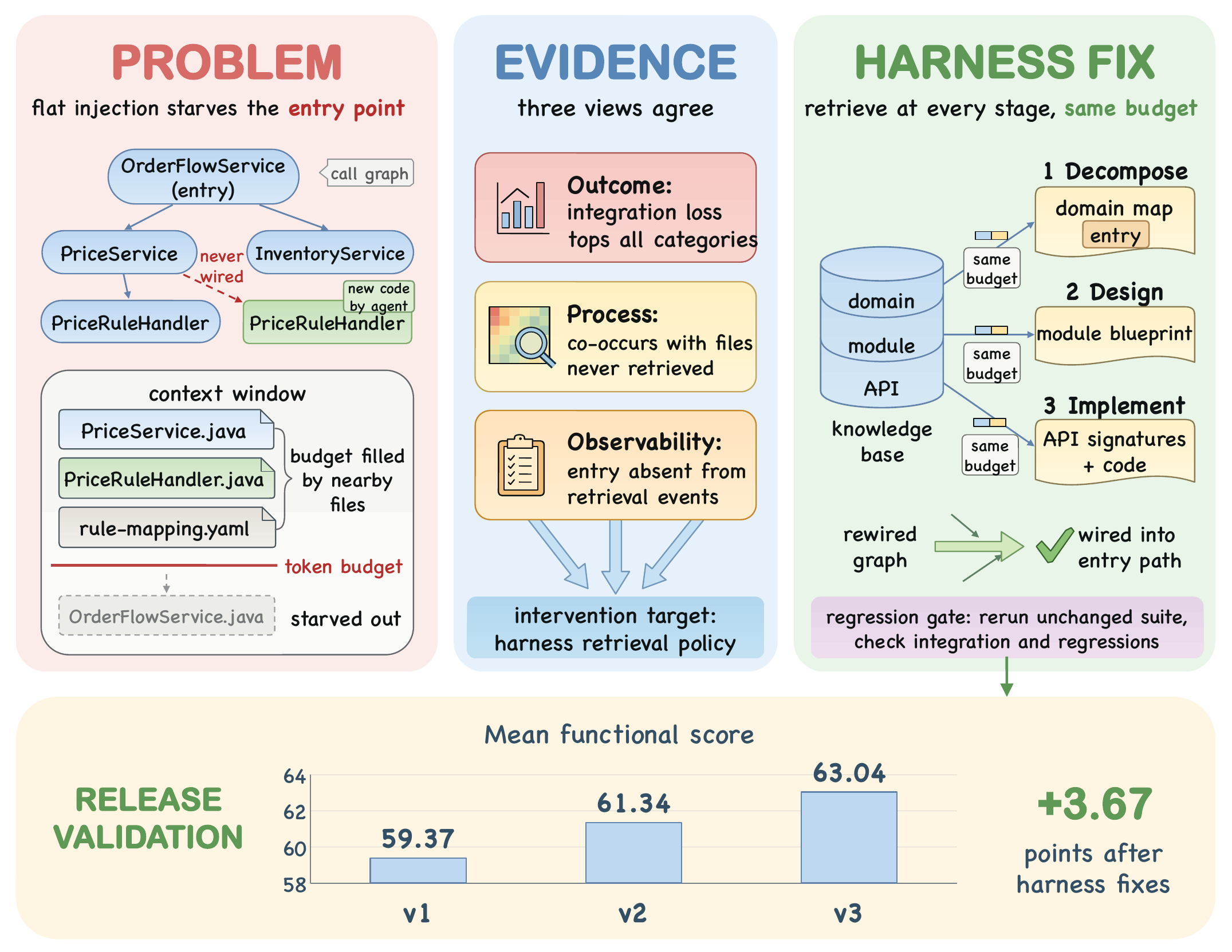}
  \caption{A representative retrieval diagnosis and scores of three successive harness releases.}
  \label{fig:deployment-case}
\end{figure}

Figure~\ref{fig:deployment-case} illustrates one retrieval repair.
Acceptance gaps, unretrieved entry files, and retrieval events linked
missing integration to context selection. Stage-matched retrieval
replaced flat injection with domain, module, and API context at the
appropriate task stage under the same budget. This changes what the
agent sees at implementation while keeping the context allowance
fixed, so the same acceptance conditions can be evaluated after the
change.

\begin{figure}[t]
  \centering
  \includegraphics[width=\columnwidth]{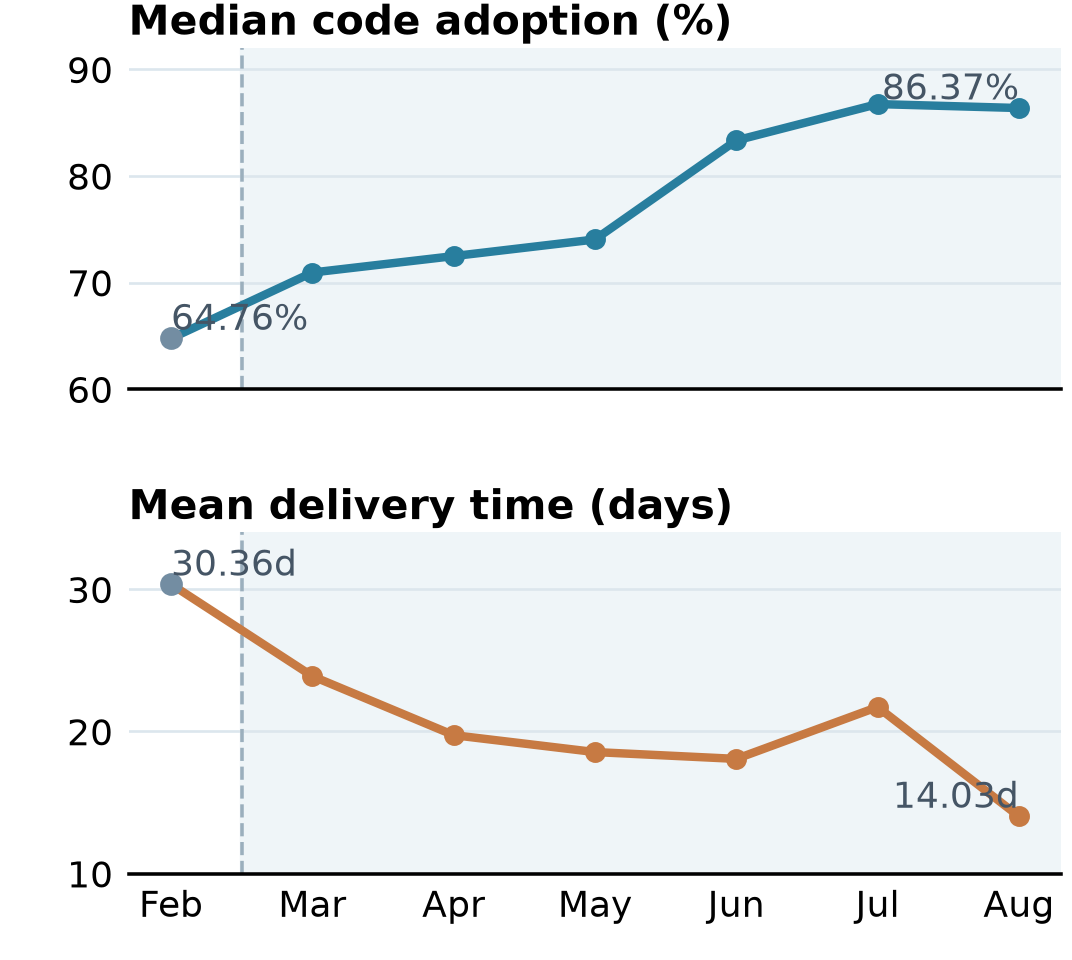}
  \caption{Monthly coding-agent platform observations from February to August.}
  \label{fig:platform-trend}
\end{figure}

\pace{} turns diagnosis into continuing harness evolution. Engineers
inspect weighted acceptance gaps and trace signals, update retrieval,
tool guidance, and planning behavior, then rerun the unchanged 43-task
suite. Across three Opus-4.5 releases, mean score rises from 59.37 to
61.34 to 63.04 (\textbf{+3.67 points};
Appendix~\ref{app:harness-releases}). During the platform rollout of
this benchmark-guided iteration, monthly median code adoption rose
from 64.76\% to 86.37\%, while mean requirement delivery time fell
from 30.36 to 14.03 days between February and August
(Figure~\ref{fig:platform-trend}).

\section{Related Work}
\label{sec:related}
\paragraph{Tasks and acceptance oracles.}
SWE-bench and SWE-bench Pro study issue resolution in existing repositories
\citep{jimenez2024swebench,deng2025swebenchpro}; FEA-Bench and
FeatureBench extend this setting to feature requests
\citep{li2025feabench,zhou2026featurebench}. DevEval and PRDBench use
PRD-style inputs for new projects
\citep{li2025devevalfull,fu2026prdbench}, while Harvest curates production
prompts with test-based scoring \citep{jha2026reap}. Table~\ref{tab:benchmark-position} distinguishes \pace{} by combining
final PRDs, existing repositories, and reviewed acceptance criteria.

\paragraph{Grounded acceptance judgment.}
LLM judges are sensitive to their criteria and evidence presentation
\citep{zheng2023judging,wang2024fairevaluators,gu2024judgesurvey}.
Rubrics and checklists make the target explicit
\citep{liu2023geval,lee2025checkeval,starace2025paperbench}; CodeJudge
scores code deviations \citep{tong2024codejudge}, and Agent-as-a-Judge
searches the workspace \citep{zhuge2024agentjudge}. \pace{} fixes
expert-audited Code Point contracts before evaluation, localizes code
and integration evidence, and aggregates cited criterion judgments by
rule. Its controls, expert agreement, and repeated judging test this
acceptance boundary on the same frozen suite.

\paragraph{Diagnosis for agent improvement.}
Trace studies localize execution failures and analyze agent behavior
\citep{deshpande2025trail,cemri2025mast,zhang2025which}.
\pace{} connects those process signals to weighted unmet PRD conditions.
Engineers inspect the linked evidence, revise the harness, and rerun
unchanged tasks; the release comparison in Section~\ref{sec:deployment}
shows how the benchmark is used beyond model ranking.

\Needspace{12\baselineskip}
\section{Conclusion}
\pace{} provides a benchmark and evaluation methodology for enterprise
PRD development. Incomplete implementation and missing integration
emerge as the principal bottlenecks. Expert agreement and repeated
judging support evaluator reliability: judging varies substantially
less than agent execution. Similar functional means nevertheless
conceal differences in delivered requirements, motivating analysis
beyond aggregate scores.

Acceptance losses, execution signals, and quality observations guide
harness refinement and provide a basis for extending this feedback to
knowledge evolution and targeted model training.

\section*{Limitations}
\paragraph{Data sources.}
Our benchmark focuses on tasks grounded in real user--agent interactions
and the corresponding delivered code. We do not investigate alternative
task construction approaches, such as mining tasks directly from merge
requests or deriving them from repository structure. These sources may
capture different types of development tasks and provide complementary
coverage.

\section*{Ethical Considerations}
The study uses internal development records under organizational data
governance policies. The conversations, repositories, and requirements
do not contain end-user or customer data. Developers were informed that
platform sessions were recorded for engineering improvement, and
personal identifiers were removed before analysis.

Code Point selection and revision involved technical experts deeply
familiar with the corresponding business domains and requirements.
The agreement study involved technical experts experienced in frontend
and backend evaluation and coding agents. Reviewers were company
employees based in China, informed of the review tasks, and agreed to
participate as part of their regular professional duties.

\paragraph{Potential risks.}
Development records may contain identifying information in conversation
text or metadata. Identifier removal and organizational access controls
reduce the associated privacy risks.

\bibliography{references}

\appendix
\section{Experimental Configuration}
\label{app:config}
The primary common-scaffold study uses OpenHands 1.21.0 with a terminal
and patch tools. Native-scaffold rows retain their corresponding Codex
or QwenCode configuration. Ground-truth criteria, weights, base revisions,
and candidate patch identities remain fixed during evaluation.

\begin{table}[ht]
\centering\small
\begin{tabular*}{\columnwidth}{@{\extracolsep{\fill}}p{.43\columnwidth}p{.47\columnwidth}@{}}
\toprule
Item & Configuration \\
\midrule
Cases / Code Points & 43 / 750 \\
Common scaffold & OpenHands 1.21.0 \\
Agent iteration limit & 200 \\
Tool concurrency & 4 \\
Qwen generation effort & xhigh \\
Semantic judge & Opus-4.8 \\
Judge reasoning setting & Model default \\
Maximum judge output & 8,192 tokens \\
Point verdicts & Covered / Partial / Missing \\
Success criterion & Defined in Section~\ref{sec:judgment} \\
\bottomrule
\end{tabular*}
\caption{Common evaluation settings; native-scaffold configurations are recorded separately.}
\label{tab:config}
\end{table}

\paragraph{Evaluation scope.}
Every main-table configuration covers the same 43 cases and all
750 Code Points. Acceptance criteria, point weights, base revisions,
and candidate patches remain fixed for each comparison. Case scores
use the complete set of acceptance verdicts, and system scores give
equal weight to each case.
Table~\ref{tab:knowledge} compares the 28 cases (525 Code Points) common
to its four arms. Both studies use frozen acceptance criteria and weights,
but their judgment records differ. The main-table Qwen result retains
earlier point judgments outside its targeted review; the knowledge
comparison begins with new judgments on distinct patches before applying
the fixed targeted review. Knowledge effects are therefore paired within
Table~\ref{tab:knowledge}; its no-knowledge mean is not the 43-case
main-table mean.

The Qwen3.8-27B row uses 43 frozen final patches under
the same evaluation scope. Seventeen patches are empty. Its generation
context and recovery configuration are retained separately from the
primary eight-model study; it is not included in the diagnostic and
repetition aggregates.

\paragraph{Result validation.}
Validation checks the exact case and point sets, source patch hashes,
judge identity, evaluation configuration, and finite scores. Main-table
records contain all 750 Code Points; the knowledge comparison contains
the 525 points in its common subset. All accepted verdicts are Covered,
Partial, or Missing.
Operational failures are retained separately from accepted scoring
records and do not become implementation-failure labels.

\section{PRD Reconstruction Protocol}
\label{app:prd-protocol}
Matched requirement documents and the user's task request establish
scope. Later user clarifications refine that scope and supersede earlier
statements where the change is explicit. Task titles and changed-file
metadata validate alignment but do not introduce requirements. The
reconstructed PRD expresses the final requested behavior as work to be
performed from the base repository, avoiding conversational progress
statements. Its sections cover goals, scope, behavior, business rules,
technical constraints, and acceptance criteria where supported.
Appendix~\ref{app:prd-prompt} gives the reconstruction prompt.

\subsection{Sources of the Final PRDs}
\label{app:prd-sources}
The 43 final PRDs follow three mutually exclusive source paths
(Table~\ref{tab:prd-sources}). A case enters the clarification group when
later user instructions contribute to its final PRD, even if an initial
task document is also available. Documents may be directly supplied or
reconstructed from the production record; document provenance does not
by itself establish when an implementation name first appeared.

\begin{table}[ht]
\centering\small
\setlength{\tabcolsep}{3pt}
\begin{tabular*}{\columnwidth}{@{\extracolsep{\fill}}>{\raggedright\arraybackslash}p{.83\columnwidth}r@{}}
\toprule
Final PRD source & Cases \\
\midrule
Direct or reconstructed task document, without later user clarification & 8 \\
First user request alone & 3 \\
Initial task plus later user clarification & 32 \\
\midrule
Total & 43 \\
\bottomrule
\end{tabular*}
\caption{Source paths contributing to the 43 final PRDs.}
\label{tab:prd-sources}
\end{table}

\section{Process Observability per Model}
\label{app:process}

Table~\ref{tab:observations} reports process metrics from the same frozen
execution records as the primary main-table configurations, in the same
within-scaffold order. Counts are averaged over 43 cases; post-edit
verification instead reports the number of cases with a successful
strict project-verification command after the last observed edit.
Native-scaffold metrics are reconstructed from recorded commands,
file edits, and matched tool outcomes using the same command classifier.
Counts reflect exported events; child-agent activity is counted only
when present in the logs. These observations do not alter model scores.

\section{Consensus Reference Protocol}
\label{app:annotation}
The agreement study uses a consensus reference formed through manual
re-evaluation of all 750 Code Point verdicts for the Qwen Run~1 patches.
Three platform-side technical annotators review the requirements,
candidate implementations, verdicts, and cited evidence. Their task is
to assess whether the assigned scores are justified by the delivered code.

This review is separate from criterion validation. Business-domain
experts check requirement grounding and whether key acceptance items are
missing from the criterion set. The two reviewer groups have no personnel
overlap. The seven-method comparison uses the same candidate patches and
consensus labels and reports 95.2\% agreement and linearly weighted
$\kappa=.945$ for the evaluated full pipeline.

\section{Agreement Metric Definitions}
\label{app:metrics}
Exact agreement is the fraction of the 750 verdicts equal to the
reference: $N^{-1}\sum_i\mathbb{1}(y_i=y_i^{\mathrm{ref}})$.
Linearly weighted Cohen's $\kappa$ discounts chance agreement and
distinguishes adjacent from opposite-extreme disagreements. Covered--Partial
and Partial--Missing receive agreement weight $0.5$; Covered--Missing
receives weight $0$. Thus exact agreement measures identical decisions,
while weighted $\kappa$ also reflects their ordinal distance.

\begin{table}[ht]
\centering\small
\begin{tabular*}{\columnwidth}{@{\extracolsep{\fill}}lrrr@{}}
\toprule
 & Missing & Partial & Covered \\
\midrule
Missing & 1 & 0.5 & 0 \\
Partial & 0.5 & 1 & 0.5 \\
Covered & 0 & 0.5 & 1 \\
\bottomrule
\end{tabular*}
\caption{Linear agreement weights for Cohen's $\kappa$.}
\label{tab:kappa-weights}
\end{table}

\section{Complete Per-Point Acceptance Procedure}
\label{app:cascade}
The point contract freezes criterion IDs and text, identifier and
location constraints, allowed implementations, and weight. Patch
reconstruction verifies the exact base, file boundaries, and artifact
identity before scoring. Input failures remain evaluation errors.

\paragraph{Structural certainty.}
L0 checks mandatory structures within their complete scope. It returns
Missing for conclusive violations and otherwise defers to L1; it does
not return Covered. Deferred cases include legal
interfaces, annotations, and delegation. A name occurrence by itself
does not establish behavior.

\paragraph{Evidence package and stopping rule.}
L1 distinguishes hard anchors from retrieval seeds. Its deterministic
positive exception is restricted to a package-structure contract: the
strategy must be \texttt{symbol\_exists}, the target must be a candidate-modified
\texttt{\_\_init\_\_.py}, and any specified package name must match. This
check establishes the requested package structure, not its behavior.
Other positive judgments require L2. The evidence package records
complete declarations, method bodies, exact candidate diffs, base-context
labels, and the concrete edges needed by each criterion. An integration
claim needs source, target, and an observed call, assignment, import/export,
registry entry, or configuration binding. An explicit empty connection
inventory records the search result without granting credit.

When a checklist identifies a specific missing declaration, the localizer
extracts the owner-qualified method and performs one expansion through
same-owner helpers, bounded to two hops. It then stops. Complete relevant
inspection with no implementation establishes Missing; an unavailable
required body or external dependency remains unresolved. Retrieval hits
and file names alone never settle a behavioral condition.

\paragraph{Criterion binding and aggregation.}
L2 sees criterion text and the evidence package, but not weights or the
reference answer. Its response must contain exactly the frozen criterion
ID set. Each entry cites evidence, records severity, and judges the same
condition with its original polarity. The aggregation code handles
hard misses, partial fulfillment, and soft gaps. A connection-related
correction requires corroborating evidence from the complete L1 inventory.
Weights apply after verdict resolution; unresolved requirements never
disappear from the case denominator.

\section{Attribution Rules}
\label{app:rules}
The outcome classifier reads the cited reasoning of each non-Covered
point in a fixed priority order: contract or target mismatch,
integration or registration gap, test-delivery gap, and incomplete
implementation. Each point contributes $w_i(1-\phi(y_i))$ to one
symptom. Process rules independently scan tool events, retrieved files,
patches, verification commands, and their outcomes. A case can receive
multiple process signals. Both classifications are deterministic
functions of the saved artifacts and introduce no additional judge call.

\section{Per-Model Symptom--Signal Co-occurrence}
\label{app:per-model}
\begin{figure*}[t]
\centering
\includegraphics[width=\textwidth]{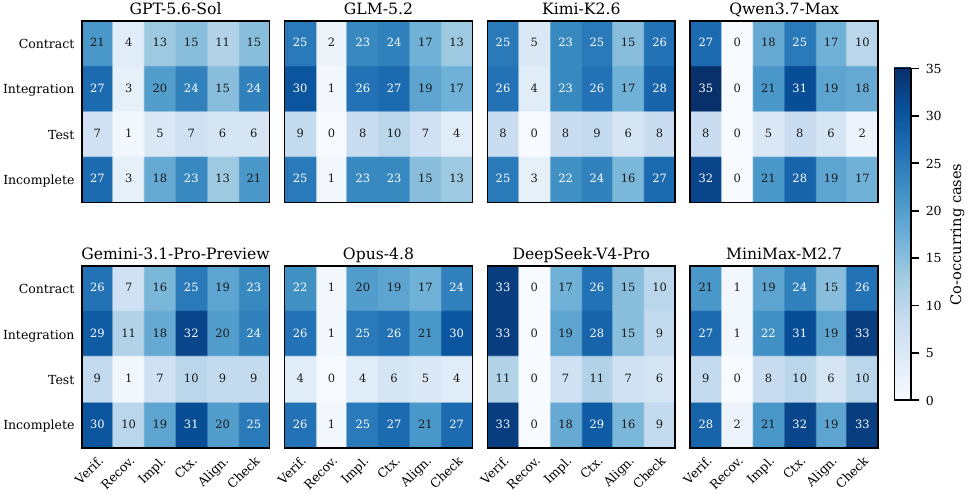}
\caption{Outcome symptoms and process signals across the eight primary models.}
\label{fig:cooccur-all}
\end{figure*}

Figure~\ref{fig:cooccur-all} extends the main three-model view to all
eight primary models with a shared color scale.

\section{Controls for Reference Grounded Contract Validation}
\label{app:controls}
Accepted reference patches supply positive controls. Each control holds
the target PRD, exact base revision, criterion definitions, and weights
fixed, changing only the candidate patch. An empty patch leaves the base
unchanged. A same-repository unrelated patch edits code outside the target
requirement. A cross-case patch transfers a change from another task;
only patches applicable to the exact target base are scored.
Identifier-only patches provide expected names or locations while
leaving the corresponding concrete classes or methods empty.

The frozen identifier-only control targets 54 behavioral criteria and
returns 51 Missing, three Partial, and no Covered. With frozen weights
summing to 109.105 and Partial weights summing to 8.8, its normalized
score is $100\times(0.5\times8.8)/109.105=4.03$.
This score summarizes the constructed control criteria, rather than a
mean over complete cases. The residual Partials reflect structural
credit for declarations without their required behavior.

The controls test whether the evaluator distinguishes required behavior
from irrelevant changes and superficial declarations. The manual
consensus comparison separately assesses judgment agreement on candidate
implementations under the same evaluation protocol.

\subsection{Positive Controls: Accepted Reference Patches}
\label{app:replay}
Reference evaluation covers 43 cases and 750 Code Points. The mean is
98.35 and the median is 100.00, with 736 Covered, 7 Partial, and
7 Missing verdicts. Business-domain experts reviewed all 14 non-Covered
points against the PRD and reference patch. They confirmed that the
criteria describe explicit requirements and that the implementation
omits the corresponding details. The Partial and Missing judgments
were therefore retained. An accepted reference is a positive example,
not a guarantee of perfect fulfillment of every PRD condition.
Reference-linked criteria may therefore receive Partial or Missing when
relevant changes omit required behavior.
Table~\ref{tab:negative-controls} reports this positive control alongside
the four negative controls.

\section{Task-Size Associations}
\begin{figure*}[t]
\centering
\includegraphics[width=.95\textwidth]{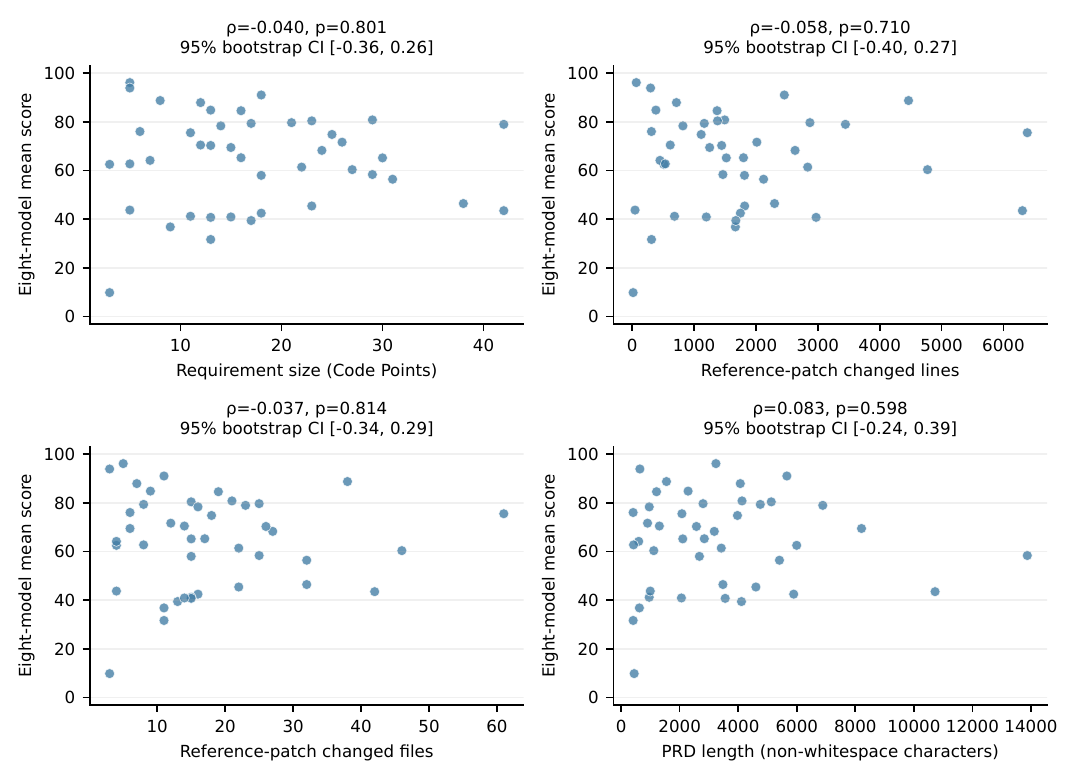}
\caption{Task-size associations with eight-model mean score. Each point is a case; intervals use case-level bootstrap sampling.}
\label{fig:appendix-size-associations}
\end{figure*}

\section{Artifact Availability}
\label{app:availability}
Code is available at \url{https://github.com/TianqiLiu1/PACE-Bench}.
The repository includes benchmark construction and evaluation code,
contract schemas, scoring logic, and attribution rules.
Production requirement documents, repository snapshots, and development
records remain subject to organizational access controls. This appendix
documents the task protocol, scoring rules, prompt renderings, and
aggregate results independently of those raw production artifacts.

\section{Harness Release Configuration}
\label{app:harness-releases}
The three release configurations in Figure~\ref{fig:deployment-case}
use the same Opus-4.5 generation model and the same 43-case suite,
comprising frontend and backend development requests. The labels
v1, v2, and v3 identify the three evaluated release configurations.
Their mean-score standard errors are 3.93, 3.92, and 3.80, respectively.
Each release completes one of 43 cases (2.3\%). Section~\ref{sec:deployment} reports
the mean scores; Table~\ref{tab:harness-changes} summarizes the changes.

\begin{table}[ht]
\centering\small
\setlength{\tabcolsep}{3pt}
\begin{tabular*}{\columnwidth}{@{\extracolsep{\fill}}l>{\raggedright\arraybackslash}p{.84\columnwidth}@{}}
\toprule
Release & Main change \\
\midrule
v1 & Frontend task flow and backend simple-task guide. \\
v2 & Fuller tool guidance, complex-task decisions, and subtask tracking. \\
v3 & Updated tool metadata, clarification rules, and plan-completion events. \\
\bottomrule
\end{tabular*}
\caption{Changes across the three evaluated harness releases.}
\label{tab:harness-changes}
\end{table}

\paragraph{Release-level changes.}
Release v1 organizes frontend work into clarification, decomposition,
design, and coding, with a backend guide focused on simple requests.
Release v2 expands tool-use guidance, adds decision-based handling of
complex backend requests, and tracks subtask completion. Release v3
updates tool metadata and backend clarification rules and introduces a
plan-completion event while retaining the coding workflow.
Table~\ref{tab:harness-changes} summarizes these differences.

\paragraph{Platform-wide improvement process.}
The platform team used acceptance losses and execution evidence to
identify harness components for inspection, then applied targeted changes
and evaluated successive releases on the same task suite. This process
covered tool guidance, complex-task handling, context selection,
clarification, and completion tracking. Figure~\ref{fig:deployment-case}
traces one such change from a missed integration requirement to a
retrieval-policy update.

\paragraph{Retrieval repair.}
The diagnosis combines an unmet integration condition, an unretrieved
entry file, and the corresponding context-selection events. Under flat
injection, nearby files exhaust the context budget before the application
entry point is retrieved. The repair organizes retrieval by task stage:
domain knowledge supports decomposition, module information supports
design, and API and implementation context supports coding. The total
context budget remains fixed, and the unchanged suite checks integration
and regressions across successive release configurations.

\paragraph{Platform rollout observations.}
Figure~\ref{fig:platform-trend} shows monthly operational metrics around
the platform rollout. February is the predeployment reference month;
March through August are post-rollout observations. Code adoption is a
monthly median and requirement delivery time is a monthly mean. Together with the fixed-suite release comparisons, these monthly
aggregates document the coding-agent platform's evolution during rollout.

\section{Extended Benchmark Analyses}
\label{app:extended}
\subsection{Repository and Reference-Change Statistics}
\label{app:benchmark-comparison}
Repository statistics cover the readable base snapshots; reference-change
statistics cover all 43 cases. Repository size counts physical lines in
recognized source files, excluding version-control metadata, dependencies,
and build outputs. Reference-change size counts additions and deletions.
These measures describe available context and change scope, rather than
assigning difficulty. Figure~\ref{fig:complexity} reports their distributions.

\paragraph{Oracle construction and validation.}
Table~\ref{tab:benchmark-position} separates criterion construction
from candidate scoring. \emph{Generated acceptance criteria} refers to synthesis of new
acceptance criteria, rather than automated mining or selection of
existing developer tests. Both PRDBench and \pace{} generate criteria
and retain human quality review. DevEval uses manually curated tests;
SWE-bench, FEA-Bench, SWE-bench Pro, FeatureBench, and Harvest recover
or select existing tests. A cross in this column does not imply that
the overall benchmark construction is manual.
\pace{} scores Code Points through
structural checks, call graph evidence localization, and semantic
checklists without executing a unit test suite as its scoring oracle.
Its generated criteria receive final expert review. This distinguishes
its combined construction and evaluation design without claiming that
automated criteria are unique to \pace{}.

Existing repo means that the candidate modifies an existing codebase.
New / edited code records added and removed lines and affected files
for patch benchmarks. DevEval instead reports complete target projects.
The PRDBench value is the median annotation scaffold size from its
Figure~4: 1,279 lines (mean 2,583.8, range 188--9,185). Its scaffold is
removed before evaluation, so this value is marked separately rather
than treated as a patch statistic. A comparable scaffold file count is
not reported. Its Table~8 gives model dependent generated and modified
line counts, not a fixed reference size. Reported mean file counts in the main table are rounded to one
decimal; n/r denotes an unreported comparable value. SWE-bench Appendix A.5 describes a median reference edit of
approximately 15 lines in one file; Harvest Table~1 reports median
changes of 39 lines across two files. The cited DevEval, FEA-Bench,
SWE-bench Pro, and FeatureBench papers provide means but no explicit
median for these edit statistics, so those rows are marked separately.
Across our 43 reference patches, the mean is 1,749.93 changed lines and
17.58 changed files, compared with medians of 1,465 and 15.

Evaluation method names identify the evidence used for scoring.
SWE-bench, SWE-bench Pro, and FeatureBench execute targeted tests and
regression checks. FEA-Bench uses unit tests, DevEval evaluates code
with unit and acceptance tests, and Harvest uses fail to pass tests.
DevEval also evaluates design artifacts separately with an LLM judge.
PRDBench executes unit tests, CLI interactions, and file comparisons
under an agent judge. PACE uses structural checks, call graph evidence,
and semantic checklists without test execution as its scoring oracle.

Input labels show the main requirement format; DevEval additionally supplies design
artifacts, FEA-Bench and FeatureBench supply signatures, and SWE-bench
Pro adds requirements and interfaces.

\subsection{Code Point Provenance and Weights}
\label{app:axes}
\label{app:annotation-details}
\label{app:weight-details}
Each point records its requirement source, implementation constraints,
and retrieval hints. Explicit source means directly stated requirement
evidence; derived and implicit points require additional interpretation
of that evidence and the accepted change. These labels describe the
point's origin, not a shortcut to acceptance. A symbolic contract can
add a structural check, while behavioral credit requires semantic judgment.
The narrow L1 package-structure exception is described in
Appendix~\ref{app:cascade}; it does not grant behavioral credit.

\paragraph{Weight construction.}
Weights reflect \textbf{verifiability and business importance}, not code
volume. The initial weight distinguishes explicit, directly checkable
contracts from criteria requiring broader semantic judgment. Business
priority then adjusts that weight to reflect the consequence of leaving
the requirement unmet. A larger patch or a greater number of changed
files does not by itself earn a larger weight.

Automated refinement separates composite outcomes, merges duplicates,
and assigns candidate weights. Final expert review freezes the per-point
values before candidate evaluation. Functional scoring and weighted-loss
attribution use these same weights. When a candidate is split into atomic
points, allocation avoids counting its parent requirement multiple times.
The initial verifiability distinction guides weighting; it does not grant
Covered without the evidence required by the acceptance protocol.

\subsection{Detailed Repetition Analysis}
\label{app:repetition-details}
\label{app:judge-stability-table}
In the repetition figure, point-level measures use the left axis and
case-level measures use the right axis. Percentage-point annotations
compare the first and third executions. Pairwise weighted $\kappa$ in
the stability table is averaged over the three judging pairs; severe
disagreement denotes Covered--Missing inversions.

The repetition analysis uses complete evaluation records over the same
43 cases and 750 Code Points. Execution variation compares three
Qwen3.7-Max deliveries. Judgment variation compares three Opus-4.8
judging records for the same Run~1 patches. Table~\ref{tab:stability}
lists their aggregate outcomes; Section~\ref{sec:stability} compares
the case-level variation.

\begin{table}[t]
\caption{Three executions and three fixed-patch judging records over the same 43 cases.}
\label{tab:stability}
\small
\resizebox{\columnwidth}{!}{%
\begin{tabular}{lr}
\toprule
Measure & Value \\
\midrule
Mean score, executions 1 / 2 / 3 & 66.70 / 65.44 / 65.37 \\
Mean score, judging records 1 / 2 / 3 & 66.70 / 66.47 / 66.40 \\
Complete deliveries, judging records 1 / 2 / 3 & 3/43 / 2/43 / 3/43 \\
Covered-point rate, judging records 1 / 2 / 3 & 53.3\% / 52.5\% / 52.4\% \\
Linearly weighted Cohen's $\kappa$ (mean) & 0.885 \\
Severe disagreement rate & 0.22\% \\
\bottomrule
\end{tabular}}
\end{table}

\paragraph{Case-level variation.}
For case $c$, we compute the sample standard deviation and the range
of its three scores. We then average each statistic over all 43 cases:
\begin{equation}
\overline{\sigma}=\frac{1}{43}\sum_c
\sqrt{\frac{1}{2}\sum_{r=1}^{3}(S_{c,r}-\bar S_c)^2}.
\end{equation}
Mean SD is 8.77 for execution and 2.36 for judging (ratio 3.72).
Mean range is 16.34 and 4.46 (ratio 3.67). The unit of this analysis
is the case, so offsetting score changes cannot disappear into a
single aggregate mean.

\paragraph{Repeated-delivery measures.}
For each prefix of the fixed Run~1--Run~3 order, pass@$k$ is the
fraction of Code Points Covered in at least one run; pass$^k$ is
the fraction Covered in every run. Case-level counterparts use
the same union and intersection operations on complete-delivery
outcomes.
The pass@$k$ and pass$^k$ notation follows \citet{chen2021humaneval}.
The bands in Figure~\ref{fig:passk} follow the nominal
binomial plotting convention; points within a case can be dependent.

\subsection{Requirement Composition and Task Size}
\label{app:difficulty-details}
The main figure groups cases by the share of derived or implicit
Code Points: $<25\%$, $25\%$--$50\%$ inclusive, and $>50\%$.
The groups contain 13, 16, and 14 cases and yield eight-model mean
scores of 74.99, 61.63, and 56.80. This groups tasks by how explicitly their acceptance conditions are
stated in the PRD; it does not assign an independent human difficulty label. The exploratory case-level association is
$\rho=-.410$ ($p=.006$; BH-adjusted $p=.076$ across twelve explored
descriptors).

Task volume supplies a complementary view. Table~\ref{tab:size-association}
and Figure~\ref{fig:appendix-size-associations} report associations
with the eight-model mean score. The intervals include zero for
all four descriptors, so these observations do not establish
independence between size and performance.

\begin{table}[t]
\centering\small
\begin{tabular*}{\columnwidth}{@{\extracolsep{\fill}}lrrr@{}}
\toprule
Descriptor & $\rho$ & $p$ & 95\% CI \\
\midrule
Code Point count & $-.040$ & .801 & $[-.36,.26]$ \\
Changed lines & $-.058$ & .710 & $[-.40,.27]$ \\
Changed files & $-.037$ & .814 & $[-.34,.29]$ \\
PRD length & $.083$ & .598 & $[-.24,.39]$ \\
\bottomrule
\end{tabular*}
\caption{Task-size correlations and case-bootstrap intervals over 43 cases.}
\label{tab:size-association}
\end{table}

\subsection{Reading Symptom--Signal Co-occurrence}
\label{app:cooccurrence-details}
Each heatmap cell counts cases with both the row's outcome symptom
and the column's process signal. Signals may co-occur within a case;
the columns are therefore not a partition. The shared color scale
supports comparison across models. Weighted-loss bars answer a
different question: how much unmet acceptance weight belongs to
each symptom. Engineers use the two views together to select traces
for inspection, then verify the suspected mechanism before changing
the harness.

\subsection{Auxiliary Observations}
\label{app:auxiliary-details}
Syntax-check scores range from 98.7 to 99.8 across models and are omitted
from the table. Test delivery is evaluated on the nine cases whose
acceptance contracts require tests. Auxiliary quality scores do not
contribute to the functional ranking.
Convention, architectural scope, naming-contract adherence, and
test-delivery observations provide context for a candidate's functional
score. Tool-use and verification records describe execution behavior.
These observations do not add acceptance credit or change the model
ranking. Keeping them separate preserves the interpretation of a
functional score as fulfillment of the frozen requirements.

Table~\ref{tab:observations} reports the four quality dimensions alongside
the execution-process observations. Syntax checks are reported here as a
range because they show little separation across models.

\subsection{Static Knowledge Construction}
\label{app:knowledge}
We organize static code and documentation into six categories:
\textbf{domain knowledge, engineering conventions, technical architecture,
APIs, module dependencies, and SDK references}. These materials describe
business concepts, module and directory organization, component
responsibilities, interface signatures and usage examples, dependency
coordinates, and SDK capabilities.

We construct three input variants. \textbf{Locators only} retains
file paths and class or method names. \textbf{Repository-matched knowledge}
additionally provides behavioral descriptions, interface information,
and related components, selected using task requirements and repository
identity. \textbf{Shared long context} combines domain knowledge and
engineering conventions with selected architecture, API, dependency,
and SDK documentation. The injected material is prepared before execution
and remains fixed throughout the task.

\section{Task Screening and Expert Review}
\label{app:curation}
The internal coding-platform team ran the first three stages
automatically. Business-domain experts joined only for final case
selection and criterion audit.

\begin{enumerate}[leftmargin=*,itemsep=4pt]
\item \textbf{Session screen (automatic).} Retain deliverable coding
requests and exclude non-coding or overly long conversations while
preserving diversity across nine business domains.
\item \textbf{PRD screen (automatic).} Use an existing or reconstructed
PRD. Exclude requirements that are too short to establish a task scope
or expose evaluation tests.
\item \textbf{Case validation (automatic).} Run independent PRD-only
and reference-diff-only Code Point passes. Reject cases with excessive
disagreement or near-duplicate requirements, then replay the accepted
reference patch. This stage retains 75 candidate cases.
\item \textbf{Expert selection and audit (manual).} Business-domain
experts who did not participate in the earlier construction select
representative cases and review requirement grounding, ambiguity,
duplicates, and missing key acceptance items. The final suite contains
43 cases and 750 expert-reviewed Code Points.
\end{enumerate}

Expert review \textbf{revised or removed 56 candidate Code Points}.
The review checks both directions: whether each proposed criterion is
supported by the requirements, and whether the criterion set omits key
acceptance items in a case. It therefore addresses coverage as well as
the validity of retained criteria. Agreement on patch verdicts is a
separate validation question and is not used as a substitute for this
requirement-coverage audit.

\paragraph{Independent review roles.}
Business-domain experts audit criterion validity. Three platform-side
technical annotators manually re-evaluate all 750 Qwen patch verdicts to assess
scoring validity. The groups have no personnel overlap. The latter study
uses the requirements, candidate patch, verdict, and cited evidence;
its consensus-reference protocol is described in Appendix~\ref{app:annotation}.

\section{Prompt Templates}
\label{app:prompts}
The following templates specify the instructions for task execution,
requirement reconstruction, Code Point synthesis, and semantic judgment.
Placeholders are filled with the corresponding task, contract, and
candidate evidence. The judge returns criterion-level evidence;
the evaluator computes the final verdict and score.

\subsection{Agent Task Instruction}
\label{app:prompt-task}
The agent receives the requirement and base repository, without Code
Points or the reference patch.
\Needspace{8\baselineskip}
\begin{promptbox}{Agent task instruction}
Implement all requirements in prd/prd.md, the final
requirement for this task.
Your current working directory is the repository root.
Run pwd first; do not assume /workspace.
Make the smallest complete production-code change that
fits the existing repository. Inspect existing patterns
before editing and run relevant tests where feasible.
Do not commit or read ground-truth artifacts.
Work directly in this repository without delegation.
Parallelize independent reads; serialize edits and tests.
Before finishing, inspect the final diff and remove
accidental or unrelated changes.
\end{promptbox}

\subsection{Intent Preserving PRD Reconstruction}
\label{app:prd-prompt}
\Needspace{8\baselineskip}
\begin{promptbox}{Requirement reconstruction}
Turn the supplied evidence into a standalone PRD.
Use matched requirement documents and the user's request
to establish scope. Preserve later user clarifications
and explicit changes to the requirement.
Use task titles and changed-file metadata only to check
alignment; they must not create new requirements.

Rewrite progress statements as work to be implemented
from the base repository. Preserve supported business
rules, interfaces, values, and boundary conditions.
Do not make a name mandatory merely because generated
code introduced it. Do not invent business context or
acceptance conditions.

Return Markdown with supported sections: goals, scope,
functional requirements, business rules, technical
constraints, acceptance criteria, and open items.
Omit sections without evidence.
\end{promptbox}

\subsection{Noise Filtering Bidirectional Synthesis}
\label{app:prompt-forward}
\Needspace{8\baselineskip}
\begin{promptbox}{Requirement view: PRD only}
Read {prd} without reference-code access.
List each independently verifiable behavior required
by the document. Separate mandatory identifiers from
implementation choices left to the developer.
Do not infer requirements beyond the supplied evidence.
Return functional_points with description, specified
identifiers, implied changes, priority, and reasoning.
\end{promptbox}

\label{app:prompt-reverse}
\Needspace{8\baselineskip}
\begin{promptbox}{Implementation view: reference diff only}
Read {file_diffs} without the requirement document.
Describe substantive changes file by file. Distinguish
new files from modifications and record relevant symbols.
Identify test files. Ignore formatting, comments, and
import-only changes. Do not invent business purpose.
Return change_points with target file, change type,
description, symbols, and test status.
\end{promptbox}

\label{app:prompt-reconcile}
\Needspace{8\baselineskip}
\begin{promptbox}{Reconciliation and automated refinement}
Compare {requirement_view} and {implementation_view}
using {requirement_summary} and {changed_file_list}.
Propose atomic acceptance conditions supported by the
requirement and evidenced in the accepted change.
Remove incidental refactoring and duplicate conditions.
Separate composite outcomes into verifiable units.

Record requirement source, retrieval hints, permitted
alternatives, and implementation constraints. For an
explicit symbolic contract, preserve its required
structure or interface value. Otherwise, names and paths
guide retrieval rather than define acceptance.
Submit candidates for final expert review.
\end{promptbox}

The reconciliation and automated audit propose candidates supported by
both views. Requirement-only findings are referred to final expert review;
reference evidence does not by itself establish complete fulfillment.

\subsection{Contract Review}
\label{app:prompt-audit}
An automated audit supplies evidence to the final technical reviewers;
it does not independently admit Code Points into the suite.
\Needspace{8\baselineskip}
\begin{promptbox}{Contract audit}
Assess each candidate using only the standardized PRD,
accepted reference patch, and proposed Code Points.
A retained condition needs requirement support and
implementation evidence. File or symbol existence alone
does not prove a required behavior.
Flag duplicates, composite conditions, unsupported names,
incidental changes, and conflicts with the requirement.
Return cited evidence and a recommendation: keep, merge,
rewrite, or remove. Mark unresolved scope questions for
expert review rather than guessing.
\end{promptbox}

\Needspace{14\baselineskip}
\subsection{Constrained Semantic Judgment}
\label{app:prompt-judge}
The complete protocol binds the checklist to the supplied criterion
identifiers and distinguishes an unmet condition from unavailable evidence.
This prompt applies to points reaching L2 after deterministic checks
and localization (Appendix~\ref{app:cascade}).

\Needspace{8\baselineskip}
\begin{promptbox}{Functional judge: fixed criterion scope}
Evaluate the supplied criteria using candidate evidence.
Preserve every criterion ID and its text. Do not add,
remove, merge, or rewrite acceptance conditions.
Judge candidate contributions; base code is context.
Accept equivalent implementations where the contract
permits them. Names alone do not establish behavior.

For every criterion, cite the declaration, behavior,
or integration edge that supports the judgment.
covered: the required behavior or structure is complete.
partial: a real contribution satisfies part of the
requirement; identify fulfilled and missing parts.
missing: the required behavior or structure is absent.
Do not treat uncertainty as partial implementation.

Preserve negative-condition polarity. For a prohibition,
absence of a violation in the inspected scope satisfies
the condition; an observed violation fails it.

If necessary code is unavailable, name the exact
owner::method or dependency in missing_evidence.
Do not infer missing implementation from unavailable
evidence. A single targeted expansion may supply it.
Return cited criterion judgments, not weights or scores.
\end{promptbox}

\Needspace{8\baselineskip}
\begin{promptbox}{Checklist schema for the complete protocol}
Input:
  description
  frozen_criteria: [{criterion_id, text}]
  candidate_evidence
  connection_inventory
  inspected_scope and evidence_omissions

Output:
  evidence_checklist: [
    {criterion_id, evidence, severity,
     judgment: covered|partial|missing}
  ]
  missing_evidence: [owner::method or dependency]

Weights and reference answers are not judge inputs.
The evaluator verifies criterion IDs and resolves
missing evidence before accepting the final checklist.
\end{promptbox}

\paragraph{Evidence expansion.}
The localizer may perform one targeted expansion of a named
owner-qualified method and same-owner helper calls for at most two hops.
After that expansion, unresolved evidence remains an evaluation-side
state. The criterion set does not change during this process.

\paragraph{Verdict folding.}
A hard Missing entry yields Missing; a partial entry or soft miss yields
Partial; all criteria fulfilled yield Covered. Integration corrections
must agree with the complete connection inventory. The complete protocol
does not assign a final case score by dropping unresolved point weight.

\Needspace{16\baselineskip}
\subsection{Auxiliary Quality Instructions}
\label{app:prompt-aux}
\Needspace{8\baselineskip}
\begin{promptbox}{Quality observations}
Compare candidate code with repository conventions and
the task's scope. Record naming, structure, error handling,
utility reuse, and style observations with code evidence.
For required tests, inspect whether delivered tests
exercise the expected behavior and assertions.
Keep these observations separate from functional scoring;
they do not replace the Code Point acceptance judgments.
\end{promptbox}

\end{document}